\documentclass[useAMS,usenatbib]{mn2e}
\usepackage{color}
\usepackage{times}
\usepackage{graphicx}
\usepackage{txfonts}
\title[A mapping study of L1174]{A mapping study of L1174 with $^{13}$CO $J=2-1$ and $^{12}$CO $J=3-2$: star formation triggered by a Herbig Ae/Be star}
\author[J. H. Yuan et al.]
{
\parbox[t]{\textwidth} {Jing-Hua Yuan$^{1}$, Yuefang Wu$^{2}$\thanks{E-mail: ywu@pku.edu.cn}, Jin Zeng Li$^{1}$, Wentao Yu$^{2}$, and Martin Miller$^{3}$}
\vspace*{6pt}\\
$^{1}$National Astronomical Observatories, Chinese Academy of Sciences, 20A Datun Road, Chaoyang District, Beijing 100012, China\\
$^{2}$Department of Astronomy, Peking University, 100871 Beijing, China; ywu@pku.edu.cn\\
$^{3}$I. Physikalisches Institut, Universit\"{a}t zu K\"{o}ln, Z\"{u}lpicher Str. 77, 50937 Cologne, Germany
}
\begin{document}
\date{Accepted 2012 October 31. Received 2012 October 27; in original form 2012 September 3}

\pagerange{\pageref{firstpage}--\pageref{lastpage}} \pubyear{2012}
\maketitle
\label{firstpage}
\begin{abstract}
   We have carried out a comprehensive study of the molecular conditions and star-forming activities in dark cloud L1174 with multi-wavelength data. Mapping observations of L1174 in $^{13}$CO $J=2-1$ and $^{12}$CO $J=3-2$ were performed using the KOSMA 3-meter telescope. Six molecular cores with masses ranging from 5 to 31 $M_\odot$ and sizes ranging from 0.17 to 0.39 pc are resolved. Large area ahead of a Herbig Be star, HD 200775, is in expanding and core 1 is with collapse signature. Large line widths of $^{13}$CO $J=2-1$ indicate the ubiquity of turbulent motions in this region. Spectra of $^{12}$CO $J=3-2$ prevalently show conspicuously asymmetric double-peaked profiles. In a large area, red-skewed profiles are detected and suggestive of a scenario of global expansion. There is a large cavity around the Herbig Be star HD 200775, the brightest star in L1174. The gas around the cavity has been severely compressed by the stellar winds from HD 200775. Feedbacks from HD 200775 may have helped form the molecular cores around the cavity. Seventeen 2MASS potential young stellar objects were identified according to their 2MASS colour indices. The spatial distribution of the these 2MASS sources indicates that some of them have a triggered origin. All these suggest that feedbacks from a Herbig Ae/Be star may also have the potential to trigger star forming activities.
\end{abstract}

\section{Introduction}

	L1174 is a molecular cloud located in the constellation of Cepheus with
	low- to intermediate-mass star forming activities \citep{kun08}. As a portion
    of the L1167/L1174 complex, L1174 was reported for the first time as a dark
    nebula by \citet{lyn62}. Investigations of L1174 have been performed with
	multi-wavelengths continuum and emission lines observations. The first mapping
	observations with molecular lines were carried out by \citet{mye83a} in $^{13}$CO
	$J=1-0$ and C$^{18}$O $J=1-0$. Studies with lines from $^{12}$CO and its isotopic
	variants have been performed by \citet{mye88}, \citet{wu92}, \citet{but95},
	\citet{bon96}, \citet{wil98}, \citet{buc02}, and \citet{wal04}.
	Wing emissions of $^{12}$CO $J=1-0$ were detected in L1174 and interpreted as
	evidence of outflow activities by \citet{wu92}.  With observations in NH$_3$
	$(J,K)=(1,1)$, \citet{goo93} detected an averaged velocity gradient of $0.87\pm0.32$
	km s$^{-1}$ pc$^{-1}$ in L1174. This region has been studied with lines from
	other molecules  with mapping and point observations, e.g., HCN and HNC
	by \citet{har89} and \citet{tur97}, C$_3$H$_3$ by \citet{mad89}, CS by
	\citet{zho89}, HC$_3$N by \citet{ful93}, DCO$^+$ and H$^{13}$CO$^+$ by \citet{but95}
	and \citet{wil98}, N$_2$H$^+$ and C$_3$H$_2$ by \citet{ben98}, and CH$_3$OH,
	c-C$_3$H$_2$ by \citet{buc02}. Efforts to search for water masers in L1174
	obtained negative results \citep{per94,fur03,sun07}.
	
	Located in L1174, a well known reflection nebula NGC 7023 has been extensively
	studied. NGC 7023 is illuminated by
	a Herbig Be star HD 200775 \citep{kun08}. There are 3 photodissociation
	regions (PDRs) in NGC 7023 located to the east, south and northwest of the exciting
	star. During the past decade, chemical and physical conditions of these PDRs have
	been investigated with observations at wavelengths from mid-infrared to millimeter
	 \citep{an03,sel07,ber08,sel10,boe10,abe10,job10,hab11,ros11}. Many of these studies
	were dedicated to spectroscopic observations of PAH emission features in this region.
	The first confirmedly detected fullerene C$_{60}$ in interstellar space was
	discovered in NGC 7023 \citep{wer04,sel07,sel10}. With observations using SPIRE
	and PACS onboard {\it Herschel}, \citet{abe10} for the first time detected
	far-infrared/submm filaments in the PDRs of NGC 7023.
	
	HD 200775 is a massive Herbig Be star with a spectral type of B3($\pm1$)e and
	a total luminosity of 15,000 $L_\odot$ \citep{her04}. The binarity of this source
	was confirmed by \citet{mon06}. \citet{ale08} estimated the masses of the primary
	and the secondary stars to be 10.7 and 9.3 $M_\odot$ respectively, and the age
	of the primary to be about $10^5$ yr. At the near vicinity, a biconical cavity has
	been very likely excavated by an energetic bipolar outflow in an earlier
	evolutionary stage \citep{fue98}. In a recent work, \citet{oka09} presented the
	detection of a flared disk around HD 200775.
	
	The distance to L1174 from the Sun has not been firmly determined. In the
	literature, several values have been quoted, e.g., 440 pc \citep{wu92,goo93,bon96},
	$429_{-90}^{+159}$ pc \citep{van98}, and 288 pc \citep{str92}. In this paper, we
	will follow \citet{wu92} to adopt the distance of 440 pc.

    In this paper, we will present a mapping study of L1174 with $^{13}$CO $J=2-1$ and
    $^{12}$CO $J=3-2$. The physical conditions and kinematics are profoundly discussed.
    With combination of millimeter and infrared data, star forming activities in L1174
    are well investigated. This paper is arranged as follows: we present a description
    of the observations and archival data in Section \ref{data}, while the results and some
    preliminary analysis are presented in Section \ref{results}; in Section \ref{discussions},
    we try to comprehensively
    discuss the data; the findings of this work are summarized in Section \ref{summary}.

\section{Observations and Data Acquisition}\label{data}
    \subsection{KOSMA observation}

    A region of $23^\prime\times15^\prime$ around L1174 was mapped in $^{13}$CO $J=2-1$ (220.399GHz) and $^{12}$CO $J=3-2$ (345.796GHz) using the K$\mathrm{\ddot{o}}$lner Observatorium f$\mathrm{\ddot{u}}$r SubMillimeter Astronomie (KOSMA) 3-m telescope\footnotemark[1] on Gornergrat near Zermatt in Switzerland. The observations were carried out with a $1^\prime\times1^\prime$ grid using the on-the-fly (OTF) mode on April 16th 2003. The reference point was at ($\alpha=21^h00^m22^s.13, \delta=68^\circ12^\prime52^{\prime\prime}.9$, J2000).

    The $^{13}$CO $J=2-1$ and $^{12}$CO $J=3-2$ lines were simultaneously observed with the dual-channel 230/345 GHz SIS receiver \citep{gra98}, whose noise temperature was about 120 K. The on board acousto optical spectrometer \citep{sch89} had 1501 and 1301 channels with bandwidths of 248 and 442 MHz at 230 and 345 GHz respectively. This resulted in channel widths of 165 and 340 kHz corresponding to velocity resolutions of 0.22 and 0.29 km s$^{-1}$ at 230 and 345 GHz. The system temperatures, $T_{sys}$, were 186 and 245 K at 230 and 345 GHz. The beam sizes at 230 and 345 GHz were determined to be  130$^{\prime\prime}$ and 80$^{\prime\prime}$ with continuum cross scans on Jupiter. The forward efficiency $F_{eff}$ was 0.93 during our observations. And the main beam efficiencies $B_{eff}$ were 0.68 and 0.72 at 230 and 345 GHz. The pointing was frequently checked on Jupiter and better than 10$^{\prime\prime}$.

    The data were reduced and visualized with CLASS and GREG programs of GILDAS software \citep{pet05}. The baselines were fitted with one order polynomial and removed for each line. The line intensities were corrected to main beam temperature scale using the formula of $T_{mb}=T^\ast_A\ F_{eff}/B_{eff}$.

    \footnotetext[1]{The KOSMA telescope has been transferred to Yangbajing, Tibet, China. }

    \subsection{Archival data}

    The \emph{Spitzer} Gould Belt Legacy Survey\footnotemark[2] \citep[program ID: 30574;][]{all07} carried out observations toward L1174 and its vicinity  with the InfraRed Array Camera \citep[IRAC;][]{faz04} and the Multiband Imaging Photometer for Spitzer \citep[MIPS;][]{rie04} onboard the \emph{Spitzer Space Telescope} \citep{wer04} in 2006 November and 2007 February. These data are publicly available on IRSA\footnotemark[3], where we retrieved the post-BCD (post Basic Calibrated Data) mosaics. They are directly used in the qualitative analysis of the distribution of dust and its spatial relation with molecular gas.

    \footnotetext[2]{http://www.cfa.harvard.edu/gouldbelt/}
    \footnotetext[3]{NASA/ IPAC  Infrared Science Archive (IRSA) is part of the Infrared Processing and Analysis Center (IPAC), which is operated by the Jet Propulsion Laboratory, California Institute of Technology, under contract with the National Aeronautics and Space Administration. http://irsa.ipac.caltech.edu/}

    Archived data from the 2MASS Point Source Catalog (PSC) were also used in our work. To guarantee the reliability of the data, we adopted the source selection criteria from \citet{li05}. These include: a) each source extracted from the 2MASS PSC must have a certain detection in all $J,H$ $\&$ $Ks$ wavebands; b) only sources with a $Ks$-band signal-to-noise ratio above 15 are selected. Young stellar object candidates were identified according to their 2MASS colour indices (e.g., $[J-H]$, and $[H-Ks]$. More details, see Sect. \ref{2MASS}). The distribution of these potential YSOs and their relationship with other sources were investigated.

    Photometric data of 6 IRAS point sources were retrieved from version 2.1 of IRAS Point Source Catalog. And their color indices and infrared luminosity are quantitatively investigated. Their relations with and effects to the molecular cloud are discussed in following analysis.

    \section{Results}\label{results}

    \begin{figure*}
    \centering
    \includegraphics[width=0.85\textwidth]{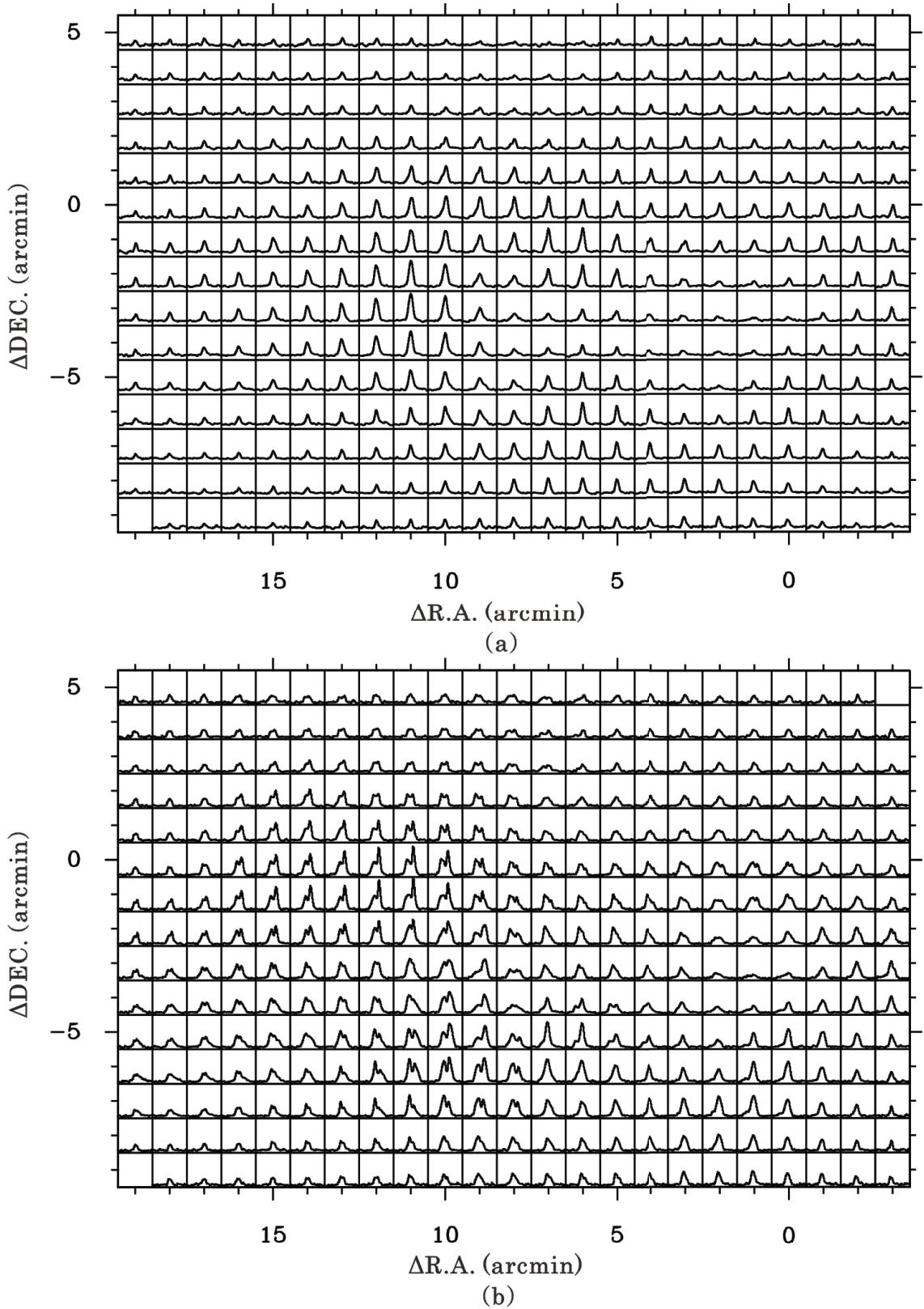}
    \caption{\label{grid}Mapping grids of L1174 and its vicinity. (\emph{a}) For the line of $^{13}$CO $J=2-1$. The velocity range for each spectrum covers -5 to 10 km s$^{-1}$, while the intensity ranges from -1 to 13 K. (\emph{b}) For the line of $^{12}$CO $J=3-2$. The velocity range for each spectrum is as the same as that in panel ({\it a}), while the intensity ranges from -1 to 6 K.}
    \end{figure*}

    \begin{figure*}
    \centering
    \includegraphics[width=0.8\textwidth]{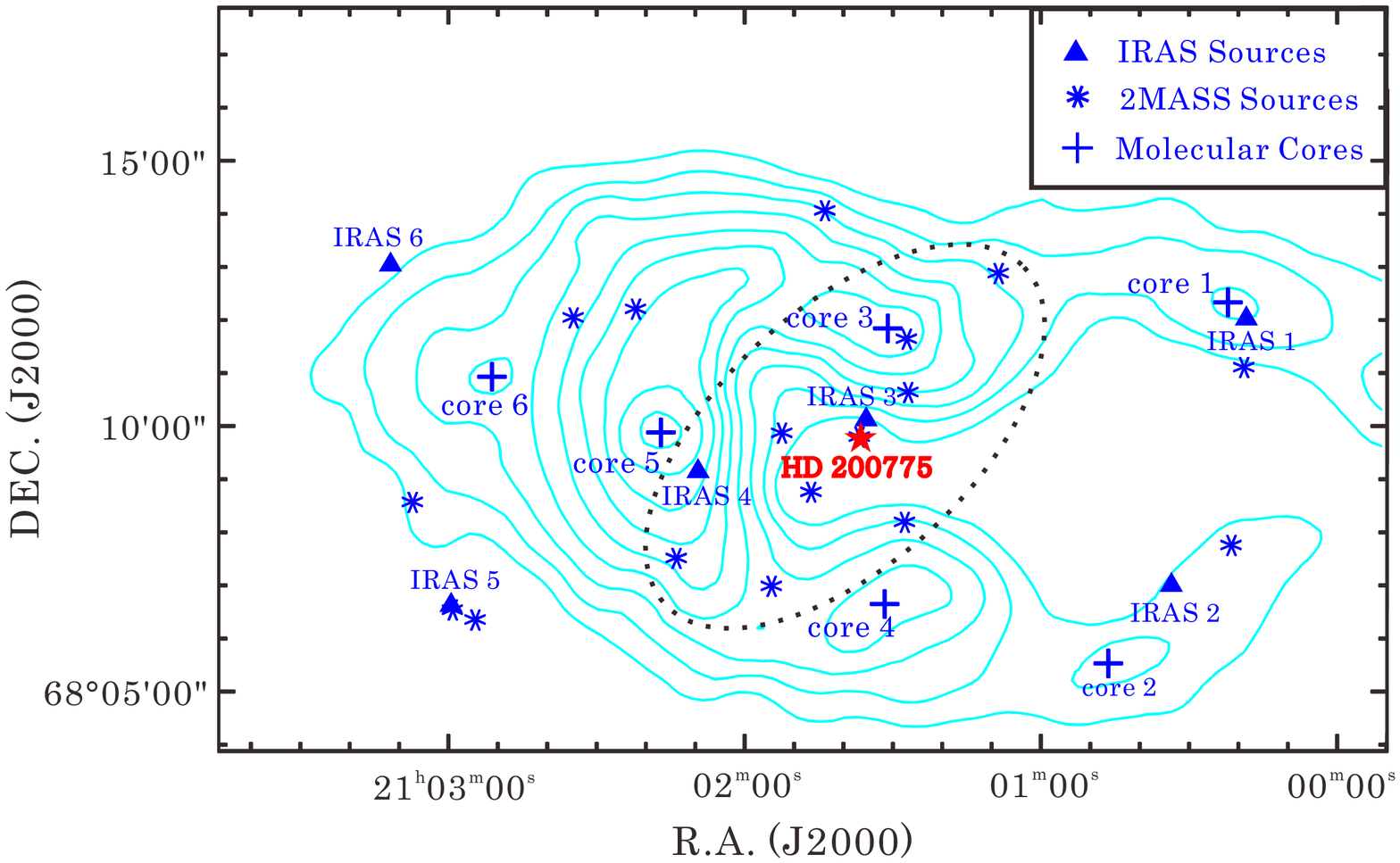}
    \caption{\label{13co}Velocity-integrated intensity map of $^{13}$CO $J=2-1$ emission. The velocity interval covers -2 km s$^{-1}$ to 7 km s$^{-1}$. The contour levels range from 4.8 to 13.1 K km s$^{-1}$ with a step of 1.5$\sigma$ (1$\sigma=0.69$ K km s$^{-1}$). The peak positions of the six molecular cores identified in this paper are labeled with crosses. The asterisks mark the candidate YSOs from 2MASS PSC, while the filled triangles indicate IRAS point sources in the field of view, whose corresponding identities are listed in Table \ref{table3}. The red star represents the Herbig Be star HD 200775.}
    \end{figure*}

    Figure \ref{grid} shows the $^{13}$CO $J=2-1$ and $^{12}$CO $J=3-2$ mapping grids of L1174 and its vicinity. $^{13}$CO $J=2-1$ spectra show mono-peak resembling gaussian profiles. In contrast, profiles of $^{12}$CO $J=3-2$ show complicated structures. This suggests that $^{12}$CO $J=3-2$ transition is optically thick in this region, while $^{13}$CO $J=2-1$ is relatively optically thin. Thus, the molecular gas traced with $^{13}$CO $J=2-1$ line is less affected by optical depth. We will use this line to derive the physical properties of the cloud.

    \subsection{Molecular cores}\label{s3.1}

    As shown in Figure \ref{grid} (\emph{a}), the intensity of $^{13}$CO line varies throughout the region. This suggests a complex distribution of molecular gas on the projected plane of sky. There exists a cavity at the mid-west. And relatively denser gas is surrounding the cavity. In the following analysis, six molecular cores with different physical conditions are resolved.

        \subsubsection{Identification}

        Figure \ref{13co} shows the velocity-integrated intensity map of $^{13}$CO $J=2-1$ rotational transition. We, by eye, identified six intensity peaks as cores, which are labeled as core 1 to core 6 with ascending right ascension order. The coordinates of the cores were converted from the offsets of intensity peaks to equatorial system (J2000) and are given in Table \ref{table1}. The uncertainties of them are referred to the pointing error, which is better than 10$^{\prime\prime}$. The radius of each core was determined according to the nearest point with $50\%$ of its peak intensity and is presented in Table \ref{table2}. We estimate that more than 80\% of the masses of the cores are involved in spheres with these radii.

        Core 1 to 5 are distributed surrounding a cavity, which has been excavated by an energetic and bipolar outflow in an earlier evolutionary stage of the massive Herbig Be star HD 200775 \citep{fue98}. And core 6 is situated at the most east, about 8$^\prime$ ($\sim1$ pc) away from HD 200775. Core 1 is associated with IRAS 20597+6800, while core 3 is coincident with a 2MASS YSO candidate (2MASS J21013520+6810086). No IRAS point sources or 2MASS YSO candidates are detected at the near vicinities of the other four molecular cores. Core 3, core 4 and core 5 show notable intensity gradients at the directions towards HD 200775. This could be attributed to the feedbacks from the central Herbig Be star.

\begin{figure}
    \centering
    \includegraphics[width=0.5\textwidth]{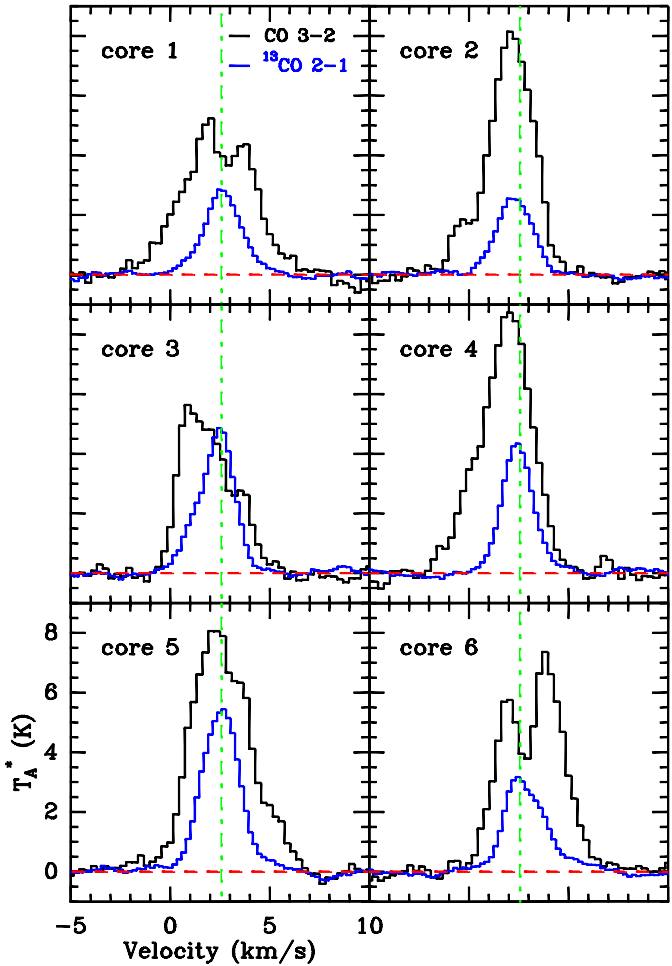}
    \caption{\label{profiles} Profiles of $^{13}$CO $J=2-1$ and $^{12}$CO $J=3-2$ emission lines at the peaks of the six cores. The red dashed line shows the zero level intensity, while the green dash-dotted line indicates the system velocity V$_{sys}=2.58$ km s$^{-1}$.}
    \end{figure}

       The spectra of $^{13}$CO $J=2-1$ and $^{12}$CO $J=3-2$ at the peak of each core are shown in Figure \ref{profiles}. Rotational line of $^{13}$CO $J=2-1$ shows similar mono-peak profiles at all six cores, while that of $^{12}$CO $J=3-2$ differs from each other. For core 1, line of $^{12}$CO $J=3-2$ indicates a blue-skewd profile  and broad line wings. This line shows mono-peak with a blue wing at core 2 and 4, while weak blue-skewed profile feature at core 3 and 5. Contrast to core 1, a red-skewed profile is observed at core 6.

        \subsubsection{LTE parameters}

        To investigate the nature of the cores, it is firstly necessary to obtain estimates of some physical parameters, e.g., temperatures, masses. The derivation of these parameters follows a rotation temperature -- column density analysis under the assumed conditions of local thermodynamic equilibrium (LTE) \citep{gar91}.

        One component gaussian fittings to molecular lines at peaks of the cores were carried out. In Table \ref{table1}, the antenna temperatures, velocity centroid, and full widths at half-maximum (FWHM) of lines of $^{13}$CO $J=2-1$ and $^{12}$CO $J=3-2$ are given.

        According to \citet{gar91}, under conditions of LTE, the measured main beam temperature $T_{mb}$ and the column density  of $^{13}$CO ($N_{^{13}CO}$) can be expressed as follows,
        \begin{equation}
        T_{mb}=\frac{h\nu}{\kappa}[\frac{1}{\mathrm{exp}(\frac{h\nu}{\kappa T_{ex} })-1}-\frac{1}{\mathrm{exp }(\frac{h\nu}{\kappa T_{bg}})-1}]\times[1-\mathrm{exp }(-\tau)]f \label{eq1}
        \end{equation}
        \begin{eqnarray}
        N_{^{13}CO}=\frac{3\kappa}{8\pi^3B\mu^2}\frac{\mathrm{exp }[hBJ(J+1)/\kappa T_{ex}]}{(J+1)} \nonumber \\
        \times\frac{(T_{ex}+hB/3\kappa)}{[1-\mathrm{exp }(-h\nu/\kappa T_{ex})]}\int\tau_{13} dv \label{eq2}
        \end{eqnarray}
        where $T_{ex}$ is the exciting temperature which we estimate to be about $20\pm4$ K from the averaged dust temperature of the 6 IRAS sources (see Table \ref{table3}), $T_{bg}=2.75$ K is the temperature of the cosmic background radiation, $f$ is the beam-filling factor which is 1 in our observations, $B$ and $\mu$ are the rotational constant and permanent dipole moment of $^{13}$CO, $J$ is the rotational quantum number of the lower state in the observed rotational transition.

        The combination of equation \ref{eq1} and \ref{eq2} would lead to a revised expression of $N_{^{13}CO}$ as,
        \begin{eqnarray}
        N_{^{13}CO}=\frac{3\kappa^2}{16\pi^3\mu^2hB^2}\frac{\mathrm{exp }[hBJ(J+1)/\kappa T_{ex}]}{(J+1)^2}\times\frac{(T_{ex}+hB/3\kappa)}{[1-\mathrm{exp }(\frac{-h\nu}{\kappa T_{ex}})]} \nonumber \\
        \times\frac{1}{[\frac{1}{\mathrm{exp}(h\nu/\kappa T_{ex})-1}-\frac{1}{\mathrm{exp }(h\nu/\kappa T_{bg})-1}]}\int{T_{mb}}\frac{\tau dv}{1-exp(-\tau)}
        \end{eqnarray}
        where $\tau$ is the optical depth of $^{13}$CO $J=2-1$. Given the optically thin feature of $^{13}$CO $J=2-1$ emission, the term $\frac{\tau}{1-exp(-\tau)}$ is approximate 1. The velocity-integrated intensities $\int{T_{mb}}dv$ for the cores are resulted from the gaussian fitting and given in Table \ref{table2}.
        To obtain the column densities of molecular hydrogen N$_{H_2}$, we adopt a canonical [CO]/[H$_2$] abundance ratio of $\simeq10^4$, as measured in nearby molecular clouds \citep{fre87,pin08}. Given the near distance of L1174, the isotope ratio [$^{12}$C]/[$^{13}$C] in the local ISM is applied, which is about 77 \citep{wil94}. With an assumed error of 10 percent for the isotope ratio, uncertainties are counted while deriving the N$_{H_2}$. The resulted column densities of $^{13}$CO and H$_2$ are given in column 4 and 5 of Table \ref{table2}.

        The finally derived mass of each core is reached with an assumption of a homogenous sphere structure.
        \begin{equation}
          M_{LTE}=\frac{4}{3}\pi R^3n_{_{H_2}}\mu_gm(H_2)
        \end{equation}
        where $\mu_g=1.36$ is the mean atomic weight of gas, $m(H_2)$ is the mass of a hydrogen molecule, and $n_{_{H_2}}=N_{H_2}/2R$ is the number density of molecular hydrogen. The resulted LTE masses of the six cores are presented in column 7 of Table \ref{table2}. The masses vary from 4.9 to 31 M$_\odot$ with a median value of 14 M$_\odot$.

\begin{table*}
\centering
\begin{minipage}{150mm}
\caption{Fitting parameters of cores.\label{table1}}
\begin{tabular}{cccccccccc}
    \hline
           &              &               &              & $^{13}$CO (J=2-1) &           & &       & CO (J=3-2) &       \\
    \cline{4-6}\cline{8-10}
    Label  & R.A.         &    DEC.       & $T^\ast_{A}$ & $\upsilon_{LSR}$ &    FWHM      & & $T^\ast_{A}$ & $\upsilon_{LSR}$ &    FWHM     \\
           & (J2000)      &   (J2000)     &     (K)  &      (km/s)      &   (km/s)     & &     (K)    &      (km/s)    &   (km/s)    \\
    \hline
    core 1 &   21:00:22.13  &  68:12:25.9   & 2.7395   &  2.638(0.014)    & 2.234(0.036) & & 5.2106   &  2.401(0.032)    & 3.974(0.083)\\
    core 2 &   21:00:45.84  &  68:05:30.7   & 2.5911   &  2.287(0.014)    & 2.126(0.033) & & 7.9352   &  2.135(0.014)    & 2.656(0.035)\\
    core 3 &   21:01:31.11  &  68:11:46.3   & 4.5975   &  2.363(0.014)    & 2.116(0.034) & & 5.3655   &  1.724(0.021)    & 3.077(0.046)\\
    core 4 &   21:01:31.76  &  68:06:39.1   & 4.2942   &  2.439(0.009)    & 1.843(0.023) & & 8.4209   &  1.911(0.017)    & 3.049(0.040)\\
    core 5 &   21:02:17.13  &  68:09:52.9   & 5.4667   &  2.543(0.010)    & 2.197(0.024) & & 7.9722   &  2.500(0.018)    & 3.437(0.042)\\
    core 6 &   21:02:51.30  &  68:10:54.7   & 3.0554   &  2.765(0.015)    & 2.385(0.038) & & 9.3281   &  3.115(0.013)    & 3.080(0.041)\\
    \hline
\end{tabular}
\end{minipage}
\end{table*}

\begin{table*}
\centering
\begin{minipage}{140mm}
\caption{Physical Parameters of cores.\label{table2}}
\begin{tabular}{ccrcccrcc}
    \hline
    Label  & R    & $\int T_{mb}dv$ & N$_{^{13}CO}$ & N$_{H_2}$   &  n$_{(H_2)}$ &   M$_{LTE}$ &  M$_{vir}$   & P$_{ex}$/$\kappa$\\
           & (pc)   &  (K km s$^{-1}$)    & ($10^{15}$ cm$^{-2}$)   & ($10^{21}$ cm$^{-2}$) & ($10^{3}$ cm$^{-3}$)  & (M$_\odot$) &  (M$_\odot$) & ($10^{5}$ K cm$^{-3}$)\\
    \hline
    core 1 & 0.17   & 8.9(0.2)  & 4.9(0.2) & 3.8(0.5) & 3.6(1.0) & 4.9(2.2) & 174.3(28.1) & $3.2(2.5)$\\
    core 2 & 0.18   & 8.0(0.1)  & 4.4(0.2) & 3.4(0.5) & 3.0(0.8) & 5.0(2.1) & 171.2(25.6) & $2.5(1.8)$\\
    core 3 & 0.20   & 14.2(0.2) & 7.8(0.3) & 6.0(0.9) & 4.8(1.2) & 10.9(2.1)& 187.6(26.1) & $3.5(2.9)$\\
    core 4 & 0.26   & 11.5(0.1) & 6.4(0.3) & 4.8(0.7) & 3.0(0.7) & 15.0(1.8)& 183.4(19.8) & $1.5(0.7)$\\
    core 5 & 0.24   & 17.5(0.2) & 9.7(0.4) & 7.4(1.0) & 5.0(1.1) & 19.3(2.2)& 247.7(26.9) & $3.6(1.8)$\\
    core 6 & 0.39   & 10.6(0.1) & 5.9(0.3) & 4.5(0.6) & 1.9(0.4) & 31.0(2.4)& 463.0(40.3) & $1.7(0.6)$\\
    \hline
\end{tabular}
\end{minipage}
\end{table*}

        \subsubsection{Virial masses}

        To collapse for a core, the mass should be larger than its virial mass. We follow \citet{mac88} to estimate the virial masses of the cores with the assumption of constant density distributions.
        \begin{equation}
        \frac{M_{vir}}{\mathrm{M_\odot}}=210\ (\frac{\Delta v}{\mathrm{km\ s^{-1}}})^2\ (\frac{R}{\mathrm{pc}})
        \end{equation}
        where $\Delta v$ is the FWHM of the observed molecular line (see Table \ref{table1}) and $R$ the estimate of radius given in Table \ref{table2}. Column 8 of Table \ref{table2} shows the resulted virial masses ranging from 171.2 to 463 M$_\odot$ which are evidently larger than the LTE masses.

    \begin{figure*}
    \centering
    \includegraphics[width=1\textwidth]{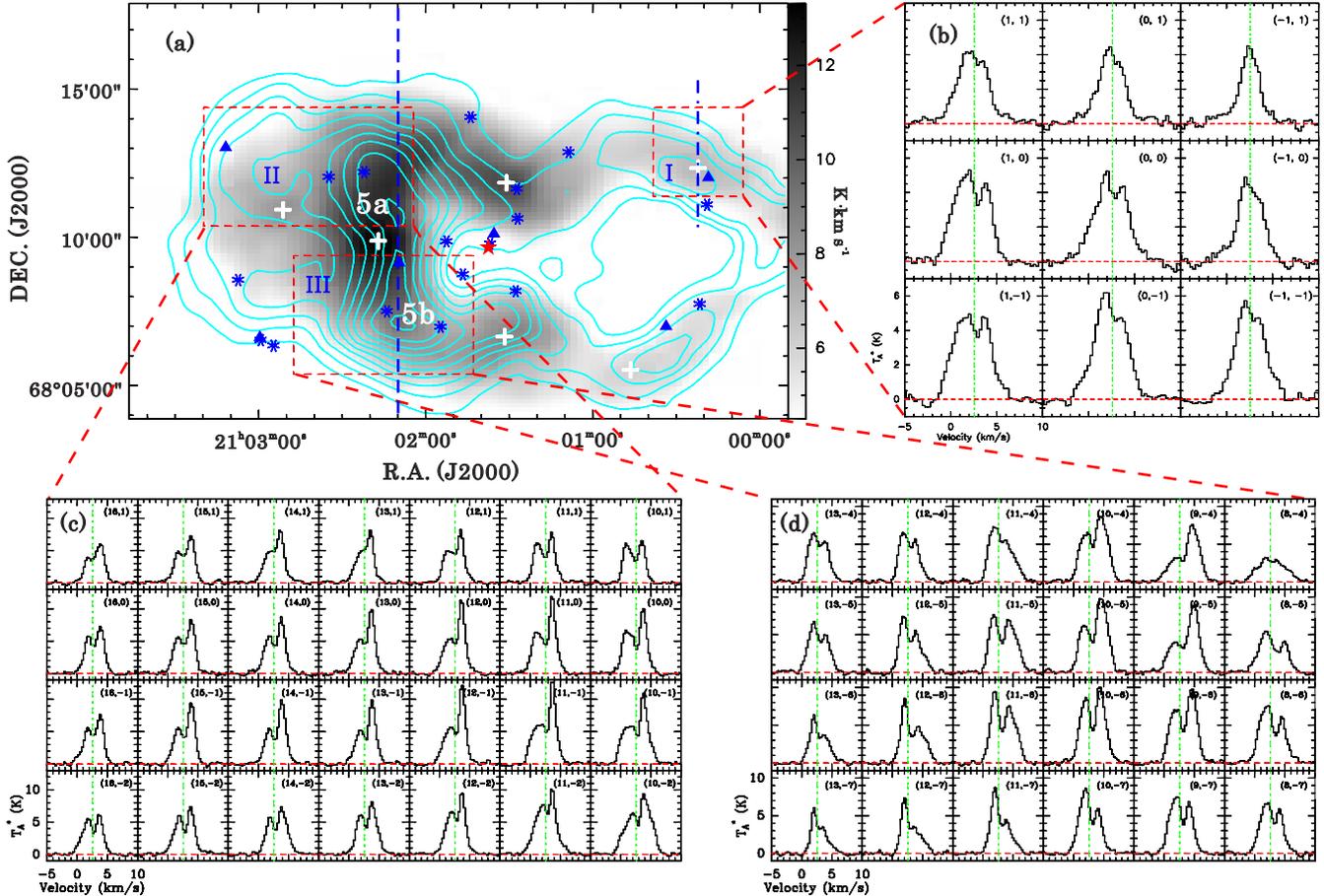}
    \caption{\label{12co} (\emph{a}) Velocity-integrated intensity map of $^{12}$CO $J=3-2$ (contours) overlayed on that of $^{13}$CO $J=2-1$ (gray). The velocity interval covers -2 km s$^{-1}$ to 7 km s$^{-1}$. The contour levels range from 14.6 to 39.3 K km s$^{-1}$ with a step of 1.5$\sigma$ (1$\sigma=1.83$ K km s$^{-1}$). The red dashed boxes locate 3 subregions with complex line profiles. Other symbols are the same as the ones in Figure \ref{13co}. (\emph{b}) Line profiles of subregion I labeled in panel (a). The offsets are indicated at the up-right corner of each sub-panel. (\emph{c}) The same as (b), but for subregion II. (\emph{d}) The same as (b), but for subregion III.}
    \end{figure*}

    \subsection{$^{12}$CO $J=3-2$ mapping}\label{12comap}

    Figure \ref{12co} (\emph{a}) shows the velocity-integrated intensity map of $^{12}$CO $J=3-2$ (contours) overlayed on that of $^{13}$CO $J=2-1$ (gray). The distribution of $^{12}$CO is more extended than that of $^{13}$CO. This is because that emissions from $^{12}$CO trace more diffuse and external part of molecular clouds. The positions of intensity peaks of $^{12}$CO are not consistent with that of cores traced by $^{13}$CO. Self-absorbtion of $^{12}$CO $J=3-2$ at densest region could explain this phenomenon. The large column density at the immediate vicinity of cores makes $^{12}$CO $J=3-2$ rotational line to be optically thick. This would cause deviation in the observed distribution of $^{12}$CO.

    Another interesting feature from the observations is the diversity of profiles of $^{12}$CO $J=3-2$ line spectra (see Figure \ref{grid} (\emph{b})). Conspicuously asymmetric double-peaked spectral lines are detected in large area of this region. Such asymmetric double-peaked profiles are always regarded as effective tracers of kinematics of molecular clouds. Asymmetrically blue profiles, lines with the peaks skewed to the blue sides, are frequently interpreted as evidence of inward motions . Such inward motions could be collapse \citep{zho93,eva05} or infall \citep{wu03,wu05}. By contrast, the red-asymmetry is observed when a centrally concentrated system (i.e., with excitation temperature decreasing outward) is expanding.

    Figure \ref{12co} (\emph{b})--(\emph{d}) show the spectral profiles of $^{12}$CO $J=3-2$ in three subregions labeled with red boxes in panel (\emph{a}). Lines in subregion \emph{I}, corresponding to core 1, show dominant blue profiles, which could be due to infall motions (see Sect. \ref{cores}). In a relatively lager area (about 1 pc in diameter), red profiles are detected in subregion \emph{II}. In our further analysis, this would be attributed to the global expansion rather than outflow (see Sect. \ref{exp}). Much more complex line profiles are observed in subregion \emph{III}, in which blue and red profiles show up alternately. The backflow of stellar winds and feedbacks from two potential 2MASS YSOs could induce complexity to the line profiles in subregion \emph{III} (see Sect. \ref{exp}).

    \subsection{Associated IRAS sources}\label{IRAS}

    \begin{table*}
    \begin{minipage}{180mm}
    \centering
    \caption{Properties of associated IRAS sources.\label{table3}}

    \begin{tabular}{ccrrrrrrrrcc}
    \hline
     &IRAS Name & $F_{12}$ & $F_{25}$ & $F_{60}$ & $F_{100}$ & $f_{qual}$\footnote{Flux density quality, 1 for an upper limit, 2 for moderate quality, and 3 for  high quality. The four numbers for each source refer to the flux density quality of IRAS bands at 12, 25, 60 and 100 $\mu$m, respectively.} & log($\frac{F_{25}}{F_{12}}$) & log($\frac{F_{60}}{F_{12}}$) & $L_{IR}$ & $T_d$\footnote{Dust temperatures were derived from color temperatures $T_{c(60/100)}$ defined by \citet{hen90}: $T_d \approx T_{c(60/100)}=\frac{96}{(3+\beta)\mathrm{ln}(\frac{100}{60})-\mathrm{ln}(\frac{F_{60}}{F_{100}})}$.} & 2MASS Association \\
      &         &    (Jy)  &    (Jy)  &    (Jy)  &    (Jy)  &             &                     &           \                &(L$_\odot$) & (K) &\\
    \hline
    IRAS 1 & IRAS 20597+6800  & 0.250     & 0.823   & 4.48     & 55.13     & 1331       & 0.52                 & 1.25                  & 7.85  & 18.96 & Null \\
    IRAS 2 & IRAS 20599+6755  & 0.677   & 0.595   & 18.98    & 1095.00      & 1231       & -0.06               & 1.45                  & 122.94 & 14.53 & Null   \\
    IRAS 3 & IRAS 21009+6758  & 26.660    & 76.770    & 637.70    & 1095.00      & 3333       & 0.46                 & 1.38                  &  364.98 & 31.02 & J21013520+6810086  \\
    IRAS 4 & IRAS 21015+6757  & 1.524    & 4.252    & 48.07    & 1095.00      & 3311       & 0.45                 & 1.50                  & 133.86 & 16.90 & Null   \\
    IRAS 5 & IRAS 21023+6754  & 0.266    & 0.387  & 0.40      & 15.46     & 3311       & 0.16                 & 0.18                  & 2.31 & 15.46 & J21025943+6806322 \\
    IRAS 6 & IRAS 21025+6801  & 0.357   & 0.796   & 10.66    & 25.17     & 3333       & 0.35  & 1.48                  & 6.46  & 28.13 & Null  \\
    \hline
    \end{tabular}
    \end{minipage}
    \end{table*}

    There are 6 IRAS point sources detected in the mapping region of KOSMA observations. They are marked in Figure \ref{13co} and Figure \ref{12co} with filled triangles and labeled as IRAS 1 to 6 with ascending right ascension. Their genuine IRAS names and IRAS photometric data from the catalog are given in Table \ref{table3}. As shown in Figure \ref{13co}, IRAS 20597+6800 (IRAS 1 in this paper) and IRAS 21009+6758 (IRAS 3 in this paper) are associated with core 1 and HD 200775 respectively.

    The infrared flux densities of the 6 IRAS sources are derived with the following equation \citep{cas86}:
    \begin{equation}
    F(10^{-13} \mathrm{W m^{-2}})=1.75\times(\frac{F_{12}}{0.79}+\frac{F_{25}}{2}+\frac{F_{60}}{3.9}+\frac{F_{100}}{9.9}),
    \end{equation}
   where $F_{12}$, $F_{25}$, $F_{60}$, and $F_{100}$ are the flux densities in Jy at 12 $\mu$m, 25 $\mu$m, 60 $\mu$m, and 100 $\mu$m respectively. With the distance of 440 pc, we obtain the infrared luminosity, which is given in column 10 of Table \ref{table3}. Color indices of log($\frac{F_{25}}{F_{12}}$) and log($\frac{F_{60}}{F_{12}}$) are calculated and shown in Table \ref{table3} either.

   Cross identification with 2MASS point source catalogue was performed for each IRAS source. The identification process followed the criteria of: a) an associated 2MASS point source must locate in the error ellipse of a IRAS source;   b) the reddest one was selected if two or more 2MASS sources were found in the very near reaches of a IRAS source. None 2MASS associations were identified for IRAS 1, IRAS 2, IRAS 4 and IRAS 6. For IRAS 3, only one 2MASS source fulfills the first criterion and was determinately identified as the near-IR association.  Two 2MASS point sources reside in the error ellipse of IRAS 5. The closer one possessing larger infrared color indices (i.e, $[J-H]$ and $[H-Ks]$) was undoubtedly identified as the association. The results of the cross identification are given in column 12 of Table \ref{table3}.

    \subsection{2MASS YSO candidates}\label{2MASS}
     \begin{figure}
    \centering
    \includegraphics[width=0.5\textwidth]{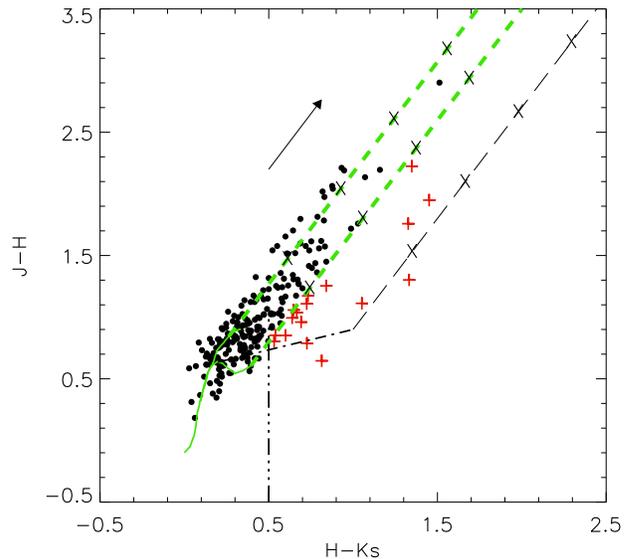}
    \caption{\label{CCD} [$J-H$] vs. [$H-K_s$] diagram. The 17 YSO candidates are marked as red "+" (see Sect. \ref{2MASS} for details). Black dots represent 2MASS field stars which may be reddened main-sequence dwarfs and giants. The solid lines are the loci of the main-sequence dwarfs and giant stars \citep{bes88}. The arrow shows a reddening vector of $A_V =5$ mag \citep{rie85}. The dot-dashed line indicates the locus of dereddened T Tauri stars \citep{mey97}. The dashed lines, which are drawn parallel to the reddening vector, define the reddening band for normal field stars and T Tauri stars. Crosses are over-plotted with an interval corresponding to $A_V =5$ mag.}
    \end{figure}

     Highly reliable 2MASS point sources were reexamined according to their colour indices (i.e., $[J-H]$, and $[H-Ks]$] to pick out the potential young stellar objects (YSOs). We dotted all the sources on the [$J-H$] vs. [$H-K_s$] diagram presented as Figure \ref{CCD}. The sources with colour indices fulfilling the relations of  $[J-H]<1.8[H-Ks]-0.103486$ and $[H-Ks]>0.5$ are regarded as YSOs \citep{li05}. This process resulted in 17 YSO candidates in the mapping region of KOSMA observations. All of them are represented with red "+" in Figure \ref{CCD}. We numbered the 17 sources according to their distances to the Herbig Be star HD 200775, i.e., 2MASS 1 as the nearest one to HD 200775 and 2MASS 17 as the farthest. In Figure \ref{13co} and \ref{12co}, they are marked with asterisks. Among the 17 YSO candidates, 3 ones were identified to be pre-main-sequence stars by \citet{kun09}, and 11 ones were identified to be YSO candidates by \citet{kir09}. Photometrical data from the 2MASS catalogue and colour indices of all YSO candidates are given in column 3 to 8 of Table \ref{table4}.

    \begin{table*}
    \begin{minipage}{135mm}
    \centering
    \caption{2MASS YSO Candidates\label{table4}}
    \begin{tabular}{lcrrrccccc}
    \hline
             \footnote[0]{$^\#$ Identified to be YSO candidate by \citet{kir09}.}
             \footnote[0]{$^\dagger$ Identified to be pre-main-sequence star by \citet{kun09}.}
             \footnote[0]{$^\ast$ 2MASS \emph{index} defined as $\alpha=\frac{[J-H]}{1.8[H-Ks]-0.1035}$, where $[J-H]$ and $[H-Ks]$ are 2MASS color indices. While $\alpha=1$, a 2MASS point source will be located on the right dashed line in Figure \ref{CCD} which is the locus dividing reddened normal field stars and T Tauri stars \citep{bes88,mey97,li05}. An younger YSO would possess a smaller 2MASS index.}
             \footnote[0]{$^\star$ Angle distance to HD 200775.}
             & 2MASS Name &   m$_J$\ \ \ &  m$_H$\ \ \ &   m$_{Ks}$\ \  &        $J-H$ &       $H-Ks$ &       $J-Ks$ &  $\alpha^\ast$ & {D$_{Angle}$}$^\star$ \\
           &            &      (mag) &      (mag) &      (mag) &      (mag) &      (mag) &      (mag) &          &    (armin)        \\
        \hline

       2MASS 1$^{\#}$ & J21013691+6809476 &    6.111 &    5.465 &    4.651 &     0.65 &     0.81 &      1.46 &       0.47 &       0.00 \\
       2MASS 2$^{\#\dagger}$ & J21012706+6810381 &    12.323 &   11.150 &  10.417 &      1.17 &     0.73 &     1.91 &       0.96 &       1.24 \\
       2MASS 3$^{\#\dagger}$ & J21014672+6808453 &    11.792 &   10.798 &    10.159 &     0.99 &     0.64 &     1.63 &      0.95 &       1.38 \\
       2MASS 4$^{\#}$ & J21015265+6809520 &    14.062 &    13.004 &    12.356 &     1.06 &      0.65 &      1.71 &      1.00 &   1.47 \\
       2MASS 5 & J21012774+6808114 &     13.999 &   13.195 &   12.663 &    0.80 &      0.53 &     1.34 &      0.94 &       1.82 \\
       2MASS 6$^{\#}$ & J21012734+6811383 &    15.378 &    13.154 &    11.806 &     2.22 &      1.35 &     3.57 &      0.96 &       2.05 \\
       2MASS 7 & J21015475+6806590 &     14.116 &     13.264 &     12.664 &       0.85 &       0.60 &       1.45 &       0.87 &       3.26 \\
       2MASS 8$^{\#}$ & J21010870+6812525 &     14.736 &     13.627 &     12.902 &       1.11 &       0.73 &       1.83 &       0.92 &       4.05 \\
       2MASS 9$^{\#}$ & J21021404+6807306 &     15.768 &     14.656 &     13.604 &       1.11 &       1.05 &       2.16 &       0.62 &       4.14 \\
      2MASS 10$^{\#}$ & J21014391+6814033 &     15.036 &     13.734 &     12.402 &       1.30 &       1.33 &       2.63 &       0.57 &       4.31 \\
      2MASS 11 & J21022228+6812121 &     16.434 &     15.180 &     14.339 &       1.25 &       0.84 &       2.10 &       0.89 &       4.86 \\
      2MASS 12 & J21023492+6812024 &     15.337 &     14.377 &     13.685 &       0.96 &       0.69 &       1.65 &       0.84 &       5.84 \\
      2MASS 13 & J21002149+6807452 &     17.019 &     15.261 &     13.934 &       1.76 &       1.33 &       3.09 &       0.77 &       7.30 \\
      2MASS 14 & J21001891+6811062 &     15.386 &     14.533 &     13.994 &       0.85 &       0.54 &       1.39 &       0.98 &       7.37 \\
      2MASS 15$^{\#}$ & J21025484+6806210 &     15.661 &     13.712 &     12.261 &       1.95 &       1.45 &       3.40 &       0.78 &       8.02 \\
      2MASS 16$^{\#\dagger}$ & J21025943+6806322 &     13.871 &     13.083 &     12.357 &       0.79 &       0.73 &       1.51 &       0.65 &       8.34 \\
      2MASS 17$^{\#}$ & J21030756+6808339 &     12.435 &     11.398 &     10.732 &       1.04 &       0.67 &       1.70 &       0.95 &       8.52 \\

      \hline
    \end{tabular}
    \end{minipage}
    \end{table*}
\section{Discussions}\label{discussions}
        \subsection{Molecular cores}
        \subsubsection{Evolutionary status}\label{cores}
        Diverse profiles of $^{12}$CO $J=3-2$ are detected in distinct cores. This, along with the distribution of the infrared sources (i.e., IRAS source and 2MASS YSO candidates), suggests deferent evolutionary status of the cores.

        \begin{figure}
        \centering
        \includegraphics[width=0.45\textwidth]{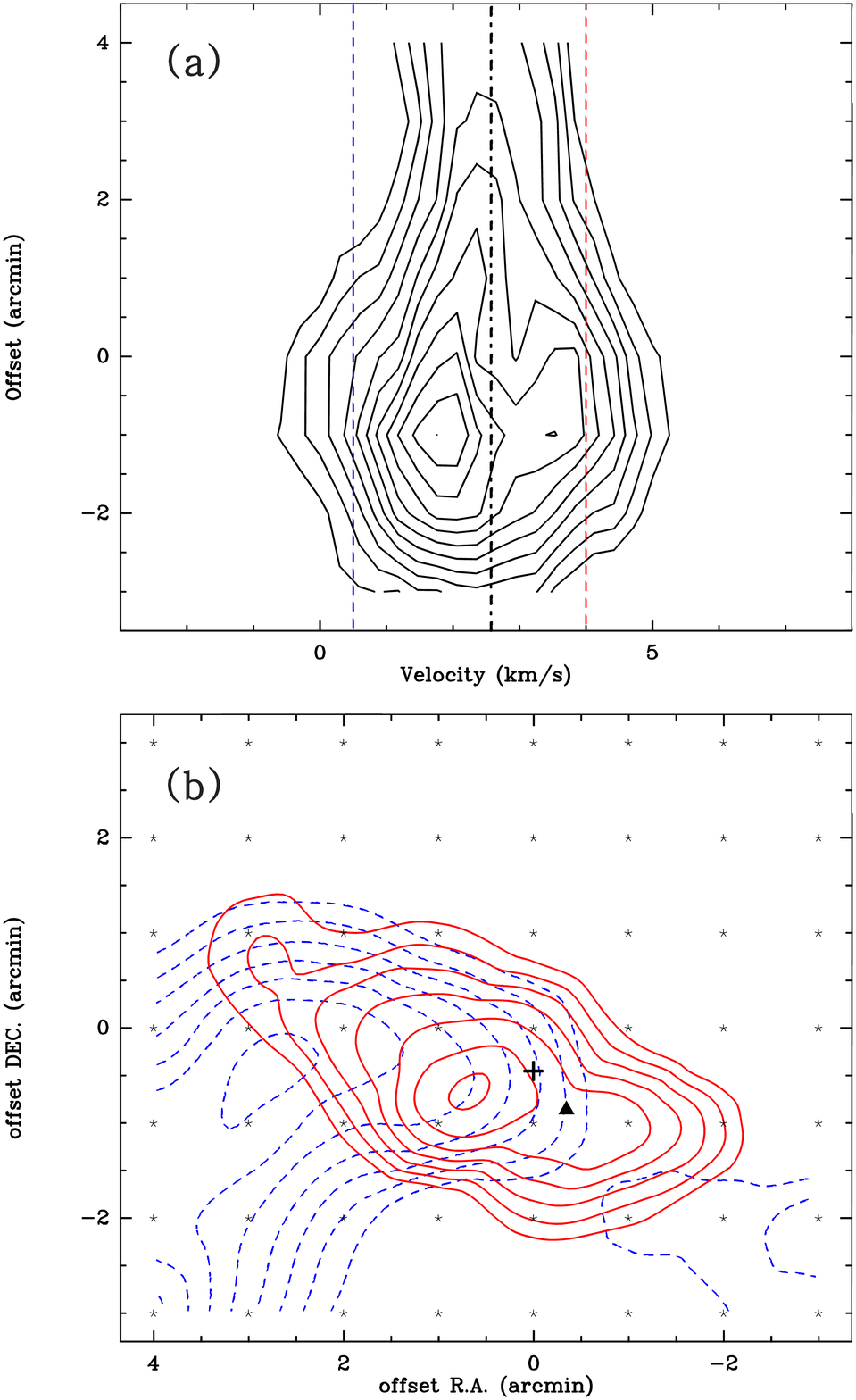}
        \caption{\label{core1-pv} \emph{(a)}$^{12}$CO $J=3-2$ position-velocity diagram constructed along the vertical dash-dotted line (\emph{cross the core 1 from north to south}) in Figure \ref{12co} \emph{(a)}. The contour levels start from the 20\% of the peak intensity with a step of 8\%. And the peak intensity is 6.18 K. The vertical dash-dotted line indicates the system velocity $V_{sys}=2.58$ km s$^{-1}$. The two dashed lines indicate the beginning of the blue and red wings. \emph{(b)} Velocity-integrated intensity map of wing emissions of $^{12}$CO $J=3-2$. The velocity interval covers -1 to 0.5 km s$^{-1}$ for the blue wing (\emph{dashed lines}), while 4 to 6 km s$^{-1}$ for the red wing (\emph{solid lines}). The contour levels start from 1.95 K km s$^{-1}$ with an increasing step of 0.37 K km s$^{-1}$ for the blue wing and start from 2.11 K km s$^{-1}$ with an increasing step of 0.33 K km s$^{-1}$. The cross presents the center of core 1, and the filled triangle indicates the location of  IRAS 20597+6800 (IRAS 1 in this paper).}
        \end{figure}

        \textbf{Core 1} is the smallest one in size. The association with IRAS 20597+6800 indicates ongoing star formation in core 1. The profile of $^{12}$CO $J=3-2$ shows blue-skewed asymmetric feature suggestive of infall motions. Based on the simple analytic model of radiative transfer proposed by \citet{mye96}, the inward speed is estimated to be about 0.43 km s$^{-1}$. Shown in Figure \ref{core1-pv} (\emph{a}) is the position-velocity (P-V) diagram  with a slice crossing the core 1 from north to south. This is a typical P-V diagram showing blue profile \citep{wu07}, which well resembles an infall scenario. Meanwhile, this P-V diagram indicates the detection of prominent line wings in core 1. Such feature is always interpreted as the result of outflow activities in star forming regions \citep{lad85}. Point observation of L1174 (corresponding to core 1 in this study) in $^{12}$CO $J=1-0$ was performed by \citet{wu92}, and the prominent wing emission was interpreted as the evidence of the existence of an outflow. And in the outflow catalog of \citet{wu96}, L1174 was classified as a source driving a bipolar outflow. Shown in Figure \ref{core1-pv} \emph{(b)} is the velocity-integrated intensity map of the wing emissions. The red lobe of the outflow is unambiguously revealed, while the blue lobe is not clear. Since core 1 is located in the large molecular envelope of HD 200775 whose feedback severely impacts the ambient material, it is difficult to separate the blue lobe of the outflow and the surrounding gas.

        \textbf{Core 2} is the lightest one in mass. The $^{12}$CO $J=3-2$ line of core 2 shows mono-peak feature. This core is relatively diffuse. No IRAS source or 2MASS YSO candidate is detected in core 2. All these are suggestive of an evolutionary stage prior to the beginning of star formation.

        \textbf{Core 3} is associated with a potential YSO 2MASS J21012734+6811383 (i.e., 2MASS 6 in this paper). This 2MASS source possesses large infrared excess ($[J-H]=2.22$, $[H-Ks]=1.35$) indicative of deeply embedded feature and an early evolutionary stage. The line of $^{12}$CO $J=3-2$ shows non-gaussian profile with the peak skewed to the blue end. This may be attributed to potential infall motions. As shown in Figure \ref{13co}, the southern portion of this core is compressed by the feedback of HD 200775.

        \textbf{Core 4} shows a mono-peak profile in $^{12}$CO $J=3-2$. The density is as low as that of core 2. The lack of infrared source in this core suggests a similar evolutionary stage as core 2.
        \begin{figure}
        \centering
        \includegraphics[width=0.45\textwidth]{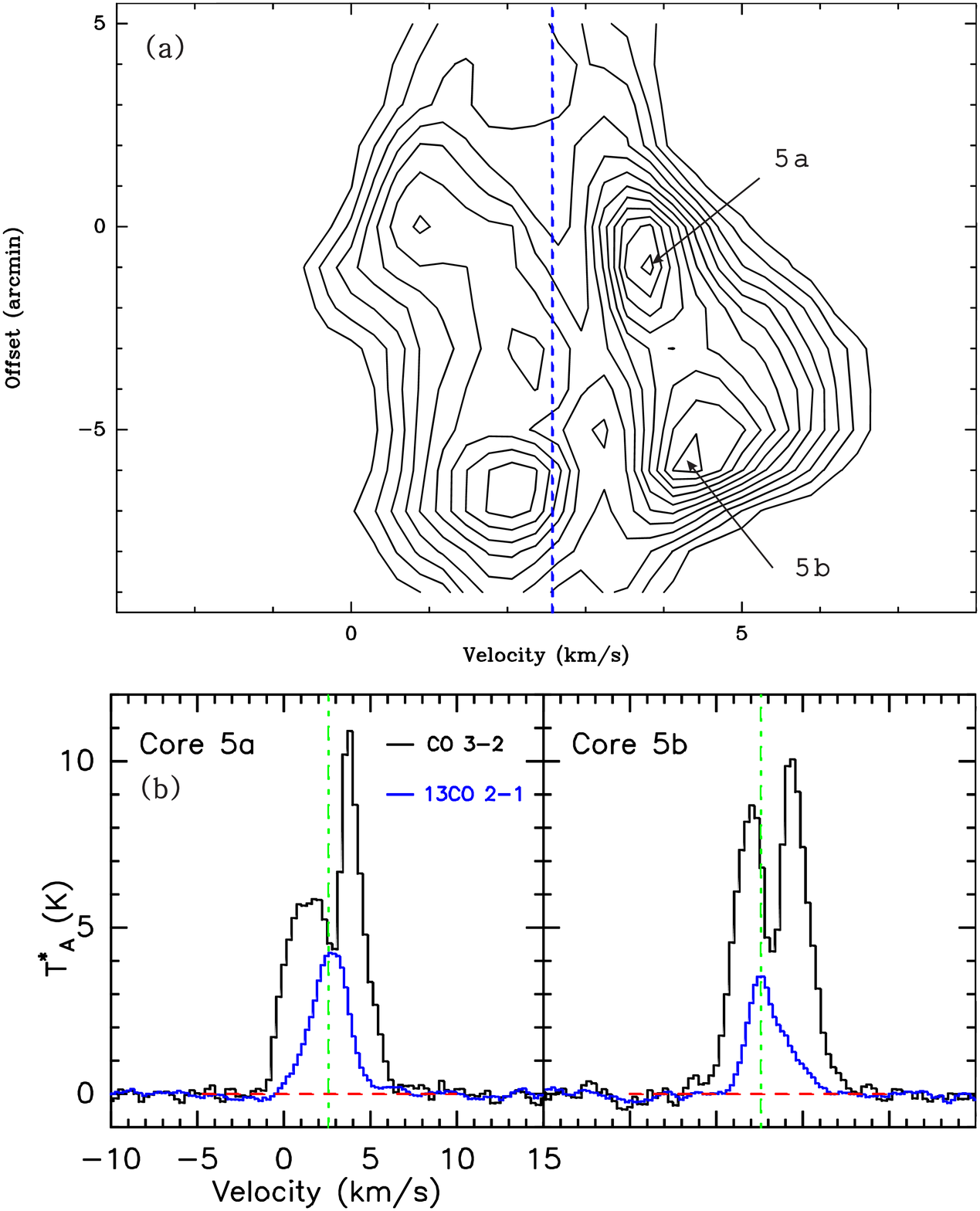}
        \caption{\label{pv} \emph{(a)} $^{12}$CO $J=3-2$ position-velocity diagram constructed along the vertical dashed line in Figure \ref{12co} \emph{(a)}. The contour levels start from the 15\% of the peak intensity with a step of 7.5\%. And the peak intensity is 10.90 K. The vertical dashed line indicates the system velocity $V_{sys}=2.58$ km s$^{-1}$. (\emph{b}) Profiles of $^{13}$CO $J=2-1$ and $^{12}$CO $J=3-2$ at the peaks of core 5a and 5b.}
        \end{figure}

        \textbf{Core 5} is the densest one. The large gradient in Figure \ref{13co} indicates that core 5 is heavily compressed by the winds from HD 200775. In the observations with $^{12}$CO $J=3-2$, this core is resolved into two cores (see Figure \ref{12co}). An IRAS source (i.e., IRAS 21015+6757) is located between core 5a and 5b. Core 5b is associated with 2MASS J21021404+6807306 (i.e., 2MASS 9 in this paper). Compared with 2MASS 6, 2MASS 9 shows smaller infrared excess ($[J-H]=1.11$, $[H-Ks]=1.05$). This indicate that core 5b could be more evolved than core 3. Shown in Figure \ref{pv} (\emph{a}) is a position-velocity diagram along centers of the core 5a and 5b. This is a typical P-V diagram showing red profile \citep{wu07}. Both of the two cores show notable red-skewed asymmetric profiles (see Figure \ref{pv} (\emph{b})). These features could be ascribed to the global expansion originated from the feedback of HD 200775 (see Sect. \ref{exp}).

        \textbf{Core 6} is the largest, most massive and diffuse one. The line of $^{12}$CO $J=3-2$ at the peak of this core shows red-skewed asymmetric profile. No infrared source is detected in this core. This, combined with the diffuse feature, suggests that this core is in a very early stage. As shown in Figure \ref{12co}, core 6 resides in the subregion \emph{II} where red-skewed lines dominate. The origin of red-skew profile of $^{12}$CO $J=3-2$ in core 6 could be due to the global expansion (See Sect. \ref{exp}).

        \subsubsection{Potential collapse}\label{collapse}

        Compared with that in other low-mass star forming regions \citep[e.g., case in Taurus,][]{oni02}, the line widths of cores in our observations are relatively large, ranging from 1.8 km s$^{-1}$ in core 4 to 2.4 km s$^{-1}$ in core 6 (FWHM of $^{13}$CO $J=2-1$ line). In regions with temperatures of about 20 K (see Table \ref{table3}), the isothermal sound speeds would be about 0.27 km s$^{-1}$ \citep{mck03}. This is much smaller than the observed line widths. Therefore, some other non-thermal motions play dominant roles in the broadening of lines in this region. Several sources could be responsible for these non-thermal line widths. One of them is that the cores formed in a turbulent environment and they are still turbulent \citep{sai06}. Meanwhile, some motions after the formation of a central protostar would lead to large line widths. Such motions could be infall, outflows, winds, and rotations \citep{pav08,wu05}. Bulk motions due to central protostar may play key roles in the case of core 1 and core 3, especially in core 1, where infall of material is detected (see Sect. \ref{cores}). Without associated infrared sources, the other four cores should be in turbulent conditions.

        From the calculation performed in Sect. \ref{s3.1}, one can see that the derived virial masses largely exceed the LTE masses for all six cores. This suggests that there isn't sufficient gravitational energy to bind the systems. \citet{lee11} and \citet{wan08} obtained similar LTE and virial masses in IRAS 05345+31571 and MWC 1080 star forming regions. \citet{sai06} also concluded that non-turbulent cores have similar virial masses to LTE masses but the virial masses are usually larger than the LTE masses for turbulent cores. External pressure is needed for a turbulent core to maintain a bound system.

        We follow \citet{sai06} to estimate the least external pressures for the $^{13}$CO cores to be in bound systems. The virial equation will be amended to be as following with external pressure:
        \begin{equation}
        F=2U+\Omega-4\pi R^3P_{ex}
        \end{equation}
        where $U=\frac{1}{2}M\sigma^2$ ($\sigma$--velocity dispersion) is the kinetic energy,  $\Omega=-\frac{3GM^2}{5R}$ is the gravitational energy, and $R$ is the radius of the core. The velocity dispersion is deduced from the line width of $^{13}$CO using the equation of $\sigma={\vartriangle}v/2\sqrt{2ln(2)}$. While $F=0$, we get the lower limits of the required external pressure. The results are then divided by Bolzmann constant $\kappa$ and tabulated in Table \ref{table2}. Notice that all the six cores need external pressures of $P_{ex}/k\sim10^5$ K cm$^{-3}$ to maintain bound systems.

        \citet{mck03} suggested that the pressure at the surface of a core is related to its surface density:
        \begin{equation}
        P_{s,core}=0.85\times10^9\Sigma^2\  \mathrm{K\ cm^{-3}}
        \end{equation}
        where $\Sigma=1.36N_{_{H_2}}m(H_2)$ is the surface density in g cm$^{-2}$. In our case, the resulted surface pressure for all the six cores is about $10^5$ K cm$^{-3}$, which is commensurate with the required external pressure for maintaining bound systems. In addition, winds from the central bright Herbig Be star HD 200775 would provide supplemental pressure to help binding the neighbour cores, i.e., core 3, core 4, and core 5. Therefore, we suggest that these turbulent cores can be bound with external pressure.

        As discussed above, turbulence plays great role in balancing the gravitational force in the cores. However, previous studies showed that this support would not exist long enough to prevent them collapsing. This could be attributed to the relatively short dissipation timescale for turbulent motions, which is about $\sim10^5$ yr, much shorter than the free-fall timescale of $\sim10^6$ yr for a core with $n_{_{H_2}}\sim10^3$ cm$^{-3}$ \citep{mck07}. This, along with the available surface pressure and feedbacks from HD 200775, suggests the six molecular cores resolved by our observations have the potential to collapse to form stars.

        \subsection{Global motions}\label{exp}

    \begin{figure*}
    \centering
    \includegraphics[width=1\textwidth]{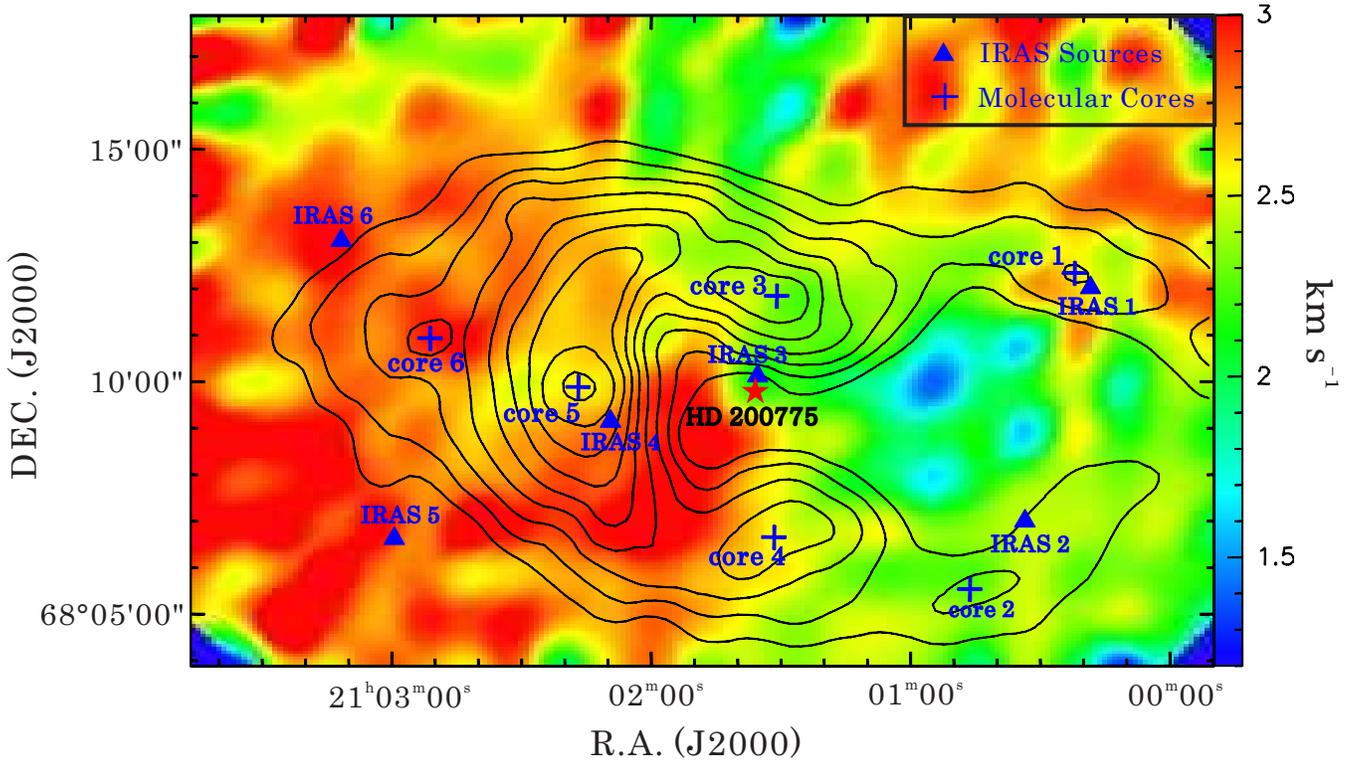}
    \caption{\label{VelDisp} Distribution of line centroid of $^{13}$CO $J=2-1$. The overlayed contours are as the ones in Fig.\ref{13co}.}
    \end{figure*}

        Comparison of spectra of subregion \emph{I} (Figure \ref{12co} {\it{(b)}}) with subregions
	\emph{II} and \emph{III} (Figure \ref{12co} {\it (c)} and \emph{(d)}) evidently shows
	variation of line profile. Obviously, the line centers for the western subregion
	 \emph{I} are slightly blue-shifted with respect to the system velocity, while
	 red-shifted for the eastern subregions \emph{II} and \emph{III}. This velocity
	 variation is also detected with $^{13}$CO $J=2-1$ line emission. The velocity
	 centroid shows a tendency of increasing with ascending right ascension (J2000) for the
	 six cores (see Table.\ref{table1}). In Figure \ref{VelDisp}, the distribution of velocity
	  centroid for the line of $^{13}$CO $J=2-1$ is presented and systematic velocity
	  variation is confirmed. The value of the velocity gradient is estimated to be about
      0.94 km s$^{-1}$ pc$^{-1}$, which is consistent with the result in \citet{goo93}.
	  This could be due to the strong stellar wind from HD 200775 which is the
	  brightest source in this region. On the other hand, this velocity gradient may suggest a
    scenario of cloud rotation with a NW-SE direction, while the southeast potion moving away from us and the
    northwest portion to us.

        Feedbacks from stars with masses $\geq$ 8 $M_{\odot}$ would profoundly affect the
	conditions of the natal clouds. The strong radiation pressure and stellar winds
	could destroy the clouds to constraint further star forming activities, otherwise
	they may help dense cores form in surroundings which will collapse to form new
	generation stars (see Sect.\ref{trigger}). In this work, HD 200775, a Herbig Be star
	with mass of $\sim10 M_\odot$, would generate violent winds blowing the whole
    molecular cloud to globally expand.
	The powerful stellar winds or outflow from this young star blew away ambient material and
	excavate a cavity at the west where the density was relatively low. The dispersed
	material accumulated around the cavity to form filamentary structures where core 1
	and core 2 have formed. For the portion to the east of  HD 200775, large amount gas
	and high density prevent it being blown about. However, the winds still possibly
	make the whole molecular to be globally expanding.

        Another observational evidence supporting a scenario of global expansion comes from the line
	profiles of $^{12}$CO $J=3-2$. As shown in Figure \ref{grid} \emph{(b)}, red-skewed
	profiles dominates in spectra with two peaks. As mentioned in Sect.\ref{12comap}, this
	is due to self-absorption of the molecular cloud with motions. Extensive red-skewed
	spectra can be observed in regions with expansion or inclined outflows. In our
	observations, the distribution of the red-skewed lines is more consistent with the
	situation of expansion. Given the fact that an outflow always has lobes constrained
    within certain directions, red-dominated profiles originated from an inclined outflow would be
    located in a region with relatively small opening angle and better collimation.
    In a scenario of an	expanding molecular cloud with a hot bright star inside, the
	blue-shifted emission originate from the hemisphere closer to the observer while
	the red-shifted emission from the hemisphere farther away; for the excitation of
	the molecules is the highest near the hot star, the observer is always looking at
	the cooler side of the blue hemisphere and the hotter side of the red hemisphere;
	thus, the red emission should always be as strong as, or stronger than, the blue
	emission (an analogue to the scenario of collapsing core proposed by \citet{zho93}).

    However, not all lines of $^{12}$CO $J=3-2$ emission resemble the red-skewed profiles.
    This ostensibly conflict with the expansion scenario. Blue-skewed spectra, rather
	than red-skewed ones, dominates at the northwest (see Figure \ref{grid} and
	\ref{12co}). This could be due to the infall motions in core 1 as aforementioned.
	The $^{12}$CO $J=3-2$ spectra detected at the southwest show mono-peak with
	relatively large wing emissions (see Figure \ref{grid} \emph{(b)}). This indicates that
	the gas column density is pretty low. Thus the line of $^{12}$CO $J=3-2$ here is
	optically thin while compared with the one in other places in this region.  Another
    confusion comes from the
	complicated spectra in subregion \emph{III} as shown in Figure \ref{grid}, where
	blue- and red-skewed line profiles are alternatively detected.
	Other mechanisms are necessary to destroy the homogeneity of $^{12}$CO $J=3-2$ line
	spectra. The blue-skewed profiles observed in the first column from right in
    Figure \ref{12co} \emph{(d)} could be due to the backflow of stellar wind.
    To the southeast of the cloud, there are two 2MASS YSO candidates, 2MASS 15 (2MASS J21025484+6806210) and
    2MASS 16 (2MASS J21025943+6806322, associated with IRAS 5). As shown in
    Figure \ref{13co} and \ref{12co} \emph{(a)},
    the cloud is as if compressed by the feedback from these infrared sources.
    With particularly large infrared excessive emission, 2MASS 15 could be in an early evolutionary
    stage with possible outflow activities which may contribute to the occurrence of the blue-skewed
    profiles in the left three columns in Figure \ref{12co} \emph{(d)}.

        \subsection{Triggered star formation}\label{trigger}

    \begin{figure}
    \centering
    \includegraphics[width=0.5\textwidth]{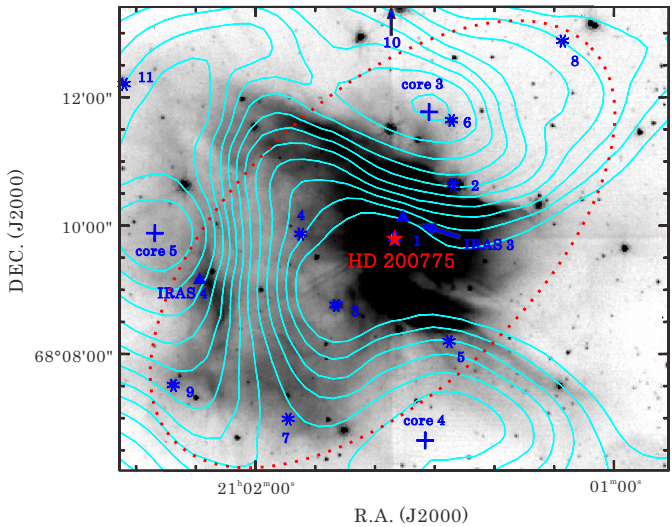}
    \caption{\label{IRAC2} Close up view of the central part of the L 11174 region. The gray scaled background with a inverted color map presents the 4.5 $\mu$m emission observed with PACS on {\it Spitzer}. The contours shows the velocity-integrated intensity of $^{13}$CO $J=2-1$. The crosses, filled triangles, and asterisks indicate the molecular cores, IRAS sources, and 2MASS YSO candidates respectively. The bright Herbig Be star HD 200775 is marked with a red star. The dotted ellipse encircles the 9 YSO candidates nearest to HD 200775. These potential YSOs are aligned in NW-SE direction.}
    \end{figure}

    \begin{figure}
    \centering
    \includegraphics[width=0.5\textwidth]{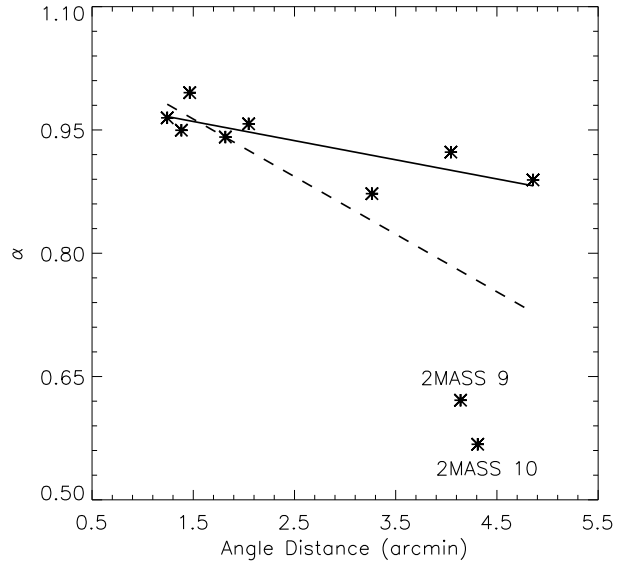}
    \caption{\label{2MASSIndex} 2MASS {\it index} as a function of the angle distance to HD 200775 for the central 10 YSO candidates.}
    \end{figure}

	Shown in Figure \ref{IRAC2} is a close up view of the central part of the region. Gray
	scaled background with an inverted color map presents 4.5 $\mu$m emission which
	mainly originate from H$_2$($v=0-0$, s(9, 10, 11)) and CO($v=1-0$) in shocked
	regions or outflows \citep{smi06,dav07,cyg08}. Evidently shocked structures are
	resolved at north and south of HD 200775. It is believed that supersonic stellar
	winds and/or outflow originated from HD 200775 have been serving as the
	engine of these shocks. Relatively weak extended 4.5 $\mu$m emission were detected
	in the east reaches. Nearby IRAS 4 (i.e., IRAS 21015+6757), an elongated structure
	still can be identified. This would be a trail feature of earlier shocks. For this
	region, we recommend a scenario of intermittent shocks generated by stellar winds
	from HD 200775.

	Overlaid contours on 4.5 $\mu$m emission in Figure \ref{IRAC2} present
	velocity-integrated intensity map of $^{13}$CO $J=2-1$ which roughly represents
	density distribution of gas. One can see large intensity gradients in sides
	toward HD 200775 for core 3, 4, and 5 which are located in sites closely
	surrounding HD 200775. These steep increases of intensities from HD 200775 to the
	near cores could be due to the strong winds from the bright Herbig Be star.
	Fierce winds have been enhancing gas density around HD 200775 to help form cores
	which would collapse to form new generation stars. If this is the case, one would
    expect to see a sharp velocity gradient around the shell excavated by HD 200775. As one
    can see in Figure \ref{VelDisp}, the velocity gradient increases some around HD 200775,
    but not sharply. This could be due to the poor spatial resolution of our observation.
    The beam of 80$\arcsec$ would smooth the velocity difference in 0.2 pc.

	Ten 2MASS YSO candidates (i.e., 2MASS 1-9 and 11) are marked with blue asterisks
	in Figure \ref{IRAC2}.
	They are labeled with numbers with increasing distances from HD 200775. Note that
	the nearest nine ones are aligned in NW-SE direction consistent with the velocity
	gradient shown in Figure \ref{VelDisp}. These features are suggestive of a scenario of
    triggered star formation in this region.
	
	We define {\it 2MASS index} for a 2MASS source as,
	\begin{equation}
	\alpha=\frac{[J-H]}{1.8[H-Ks]-0.1035}
	\end{equation}
	where $[J-H]$ and $[H-Ks]$ are 2MASS color indices. 2MASS indices for all 17 YSO
	candidates were calculated and tabled in column 9 of Table \ref{table4}.
	With an $\alpha=1$, a 2MASS
	point source will locate on the right dashed line in Figure \ref{CCD}
	which is the dividing line between reddening normal field stars and T Tauri
	stars \citep{bes88,mey97,li05}. A younger YSO would possess a smaller 2MASS
	index.

	We dotted the 10 2MASS YSO candidates nearest to HD 200775(i.e.,
	2MASS 2-11, with exclusion of 2MASS 1 which is the 2MASS association of HD 200775)
	on a plot of 2MASS index vs.
	angle distance to HD 200775 (see Figure \ref{IRAC2}). A tendency of decreasing of
	$\alpha$ with increasing angle distance is presented. To quantitatively evaluate
	the existence of such inverse correlation, we performed least square linear
	fitting. Presented as a dashed line is the fitted result with all ten points. This
	is representative a linear equation of $\alpha=(-0.070\pm0.028)d+(1.068\pm0.089)$,
	where $d$ is the angle distance to HD 200775 from a 2MASS YSO candidate. This
	indicates that the more aged YSOs, with larger 2MASS index $\alpha$, resides closer
	 to HD 200775 and younger ones
	farther away. However, a modest correlation coefficient of $R^2=0.44$ indicate that
	this tendency is not highly reliable. Shown as a solid line in Figure \ref{IRAC2},
	another fitting without considering 2MASS 9
	and 10, which possess extremely small 2MASS indices, results in a linear equation of
	$\alpha=(-0.023\pm0.007)d+(0.995\pm0.021)$ and a larger correlation coefficient of
	$R^2=0.62$. Intriguingly, the tendency of descending $\alpha$ (i.e., decreasing age)
	with increasing angle distance is better confirmed.  This feature is reminiscent of
	sequential star formation which is always interpreted as evidence of triggered star formation in the literature \citep{elm77,pre07}.

\section{Summary}\label{summary}

    We carried out mapping observations of L1174 in lines of $^{13}$CO $J=2-1$ and
    $^{12}$CO $J=3-2$ at 220.399 GHz and 345.796 GHz. Based on observed and archival data , the molecular conditions and star forming activities in L1174 were
    extensively investigated.

    Six molecular cores are resolved by the observations in $^{13}$CO $J=2-1$. Local thermodynamic equilibrium based calculations resulted in core masses of 5 to 31 $M_\odot$ with an median value of 15 $M_\odot$. Relatively large line width in all six cores indicates turbulent motions throughout the L1174 region. This also leads to large virial masses ranging from 171.2 to 463 $M_\odot$, a factor of magnitude higher than the LTE masses. However, these cores are still possible to collapse to form new stars given the available external pressure and the affects from the bright Herbig Be star HD 200775.

    Conspicuously asymmetric line profiles of $^{13}$CO $J=2-1$ spectra are observed in a large area in L1174, which indicates bulk motions in this star forming region. Purely blue-skewed line features detected in core 1 suggest the existence of infall motions. Prominent wing emissions imply potential outflows in L1174 which are not resolved because of the low resolution of our observations.

    Extensive red-skewed line features are detected to the east of HD 200775. This, together with the velocity gradient along NW-SE direction, indicates global expansion in L1174. Formed in the natal cloud, HD 200775 generated strong winds that blew away ambient material and excavated a cavity at the west where the density was relatively low. For the portion in the east, large amount of gas and high density prevent it from dispersing. And the consequence of the interaction between the winds and the gas is the global expansion of the reconstructed cloud.

    Seventeen YSO candidates were identified according to 2MASS colour indices. For the ten 2MASS YSO candidates nearest to HD 200775, a tendency of decreasing age with increasing  distance to HD 200775 is detected. This, along with large $^{13}$CO intensity gradient of cores near HD 200775 and the shock features observed at 4.5 $\mu$m, suggests a scenario of star formation triggered by the Herbig Be star HD 200775.

\section*{Acknowledgments}

     We thank the referee Paola Caselli for constructive comments that helped us improve this paper. This project is supported by the National Natural Science Foundation of China through grant NSFC 11073027, the China Ministry of Science and Technology through grants 2012CB821800 (a State Key Development Program for Basic Research) and  2010DFA02710 (by the Department of International Cooperation ).

\label{lastpage}
\end{document}